# X690→ZZ/WW/h125h95/A450Z/γγ/e+e- J=2 or J=0, narrow or wide resonance?

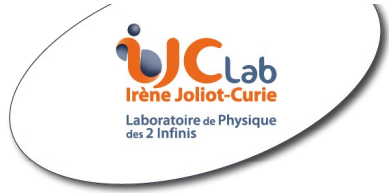

## Alain Le Yaouanc[1], François Richard[2]

Université Paris-Saclay, CNRS/IN2P3, IJCLab, 91405 Orsay, France

April 2025

## Abstract

At ICHEP 2024 CMS has published RUN2 results for ZZ→4 leptons and reached the conclusion that there is no evidence for a **scalar resonance** H650 decaying into ZZ, as was also concluded by ATLAS using an **MVA analysis** optimised for a scalar particle*. Since this resonance is indicated into **ZZ** by ATLAS in a **Cut Based Analysis** and into three other modes: **WW and h95h125 by CMS, t̄Z by ATLAS**, the present paper is an attempt to understand these apparent **contradictions**. Our conclusion is that strict selections for the reaction H650->ZZ cut as much into the signal than into the background, therefore reducing its statistical significance. A plausible interpretation of this behaviour is that these selections are inadequate for a **spin 2 particle** produced by the VBF process and decaying into ZZ with an **angular distribution similar to the Drell Yan** background. If true, our description of the various findings in term of the **Georgi Machacek model** needs to be deeply revised, which is attempted in the present note. A **scenario with T690 as a spin 2 narrow resonance** is presented which naturally accommodates WW/ZZ and h95h125 decays, while we suggest that A450Z could come from a **H650 nearby resonance**. We propose a scheme in which scalar resonances are accommodated in **three scalar isodoublets**, while T690, T450++, T375+ and T350 belong to **two tensor isoquintuplets**. In this scenario, **T690** could be a **narrow resonance which interferes with the SM background**, as confirmed by its observation in the **two photon mode**, reinforcing a **Kaluza Klein graviton interpretation**. If, as indicated by ATLAS and CMS, T690 couples to e+e-, this scenario leads to a **dramatic prediction** for future e+e- colliders.

**Presented at the 3rd ECFA workshop on e+e- Higgs/EW/Top Factories, Paris, 9-11  Oct. 24**

---

1    Alain Le Yaouanc <alain.le-yaouanc@ijclab.in2p3.fr>

2    François Richard <francois.richard@ijclab.in2p3.fr>

# Introduction

An essential input for choosing among the various options for future Higgs factories is the following: **does LHC show some convincing proof of resonances below a TeV?**

Influenced by the h125 discovery, we tend to ignore that some of these scalar candidates could well be **spin 2 resonances,** both types sharing some final states like ZZ, WW and hh, which can only be distinguished through an **angular analysis**, not yet performed in this preliminary phase. Moreover we will show that tight selections assuming a scalar resonance could be detrimental for a spin 2 particle when produced by VBF.

The **strongest candidate, X650,** is now indicated in **five final states** ZZ/WW/h95h125/A450Z with the recent [1] occurrence of X400h125→b$\bar{b}$b$\bar{b}$ also compatible with X650. Its interpretation seems incompatible with most extensions of the SM: MSSM, NMSSM and TRIPLET models (Georgi Machacek model with H$^{++}$).
Recently, CMS has updated its search for **H650→ZZ** and concluded to an absence of signal for a scalar hypothesis. We intend to show that this result and the incompatibility of X650 with available models can be interpreted as an evidence that this particle is a spin 2 particle, a **tensor**.
Before entering into this discussion, let us recall the present landscape of the various indications recorded by our work which can be summarized by the following diagram:

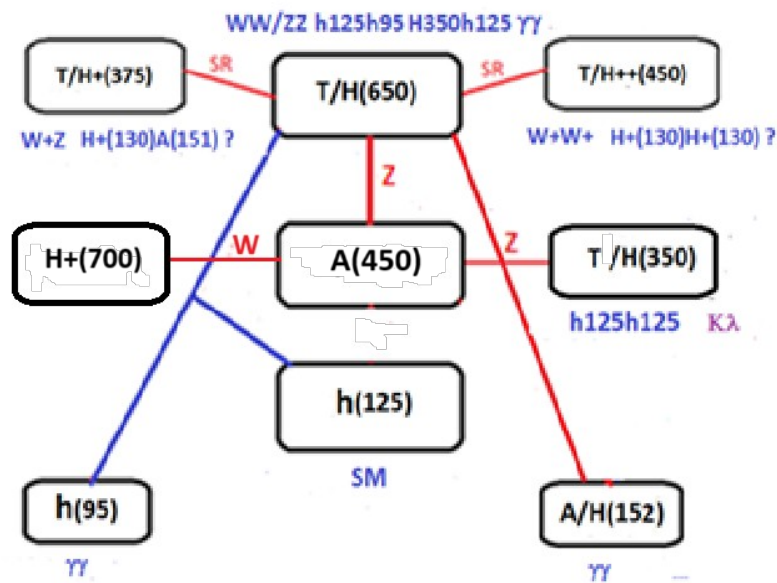

Among these channels, reaching a statistical excess above 3 s.d., only the t$\bar{t}$ narrow resonance near threshold [2] passes the 5 s.d. global significance criterion. Its favoured interpretation is a **CP-odd toponium state** ηt(345) with a cross section **7.1±0.8 pb** which agrees with prediction. A search for further scalars and pseudo-scalars decaying into top pairs was conducted without reaching a significant result.
Indications for a **CP-odd scalar A450 state** appear well above threshold in three other analyses from LHC:

- The process[3] H650→A450Z→t$\bar{t}$ℓ$^{+}$ℓ$^{-}$ [3]

- The process H$^{+}$→A450W$^{+}$→t$\bar{t}$ℓ$^{+}$ν [17]

- The process A450→h125h125Z→b$\bar{b}$b$\bar{b}$ℓ$^{+}$ℓ$^{-}$ [4]

---

3    Can also be interpreted as A650→ H450Z

Reference [5] suggests that in the case of a 2HD model, constraints from the $\rho$ parameter predict that A450 should be degenerate in mass with H650 and H$^+$ to better than 200 GeV, which is barely the case[4].

A **combination of X650 channels** allows to reach a significance above a **5 s.d. global significance**. In the X650→ZZ→$\ell^+\ell^-\ell'^+\ell'^-$ mode, this resonance looks genuinely wide, order 100 GeV, but we shall see that **interference effects** could be at work, modifying this conclusion. Its appearance as a **narrow peak** in the two photon channel discussed in IV.4 supports this hypothesis.

H650 couples to W$^+$W$^-$ as strongly as h125. Therefore the **unitarity sum rule GHW** [6] implies the existence of a doubly charged **X$^{++}$** exotic resonance with a SM-like coupling to W$^+$W$^+$. This channel has indeed been indicated by ATLAS [12]. With the coupling deduced from the SR, the measured cross section allows to extract BR(H$^{++}$→W$^+$W$^+$), of order 10%. This would imply the existence of a **light H$^+$** with large BR of X$^{++}$ into W$^+$H$^+$ and H$^+$H$^+$. ATLAS [7] has indeed a weak indication for the process t→H$^+$b with mH$^+$=130 GeV, H$^+$ decaying into $\overline{b}$c.

Finally, we will show that X650 can be interpreted as a narrow resonance, interfering with the SM background, which allows to understand its appearance in the two photon mode. For a Kaluza Klein graviton resonance interpretation one predicts BR(Gkk$^{++}$→W$^+$W$^+$)~20% (see Appendix II), also requiring a light H$^+$.

This simplified summary illustrates our approach. Starting from a solid evidence for X650, we manage to **bring together** the various other indications which, by themselves, are presently unable to reach a critical significance. Going a step further, we provide a **global interpretation** within a given model. So far, our interpretation has been based on the **Georgi Machacek model GM**. This model had to be extended to accommodate H650 with some tensions appearing in the resulting solution that showed large isospin violations and did not fulfil GM unitarity constraints.

We are also aware of other **limitations of GM,** in absence of an UV completion. As for the SM, GM could become a perturbative theory like **SUSY** or a strongly interacting **composite** model. Until this aspect is clarified, GM remains plagued by the issue of loop corrections, even more critical than for the SM, given that the $\rho$ **parameter** appears highly fine tuned[5].

Again, LHC indications suggest an alternate interpretation in terms of spin 2 resonances that will be presented in this work. Given that h125 is firmly interpreted as a scalar, this interpretation will unavoidably lead to a **combination of scalars and tensors.**

# I. The case for a spin 2 interpretation

## I.1 The ZZ channel

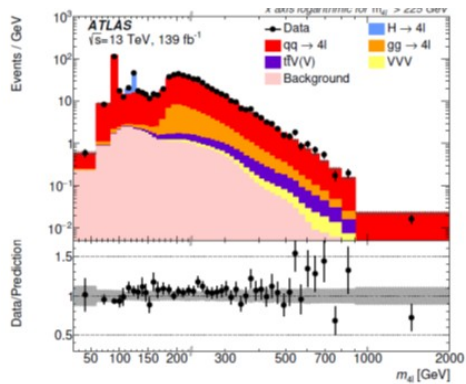

*Figure 1: The ATLAS cut based analysis [8] applied to the ZZ→4$\ell$ channel for the ggF process, indicating an excess around 650 GeV.*

---

4    Thanks to Abdelhak Djouadi for this remark
5    We are grateful to Gilbert Moultaka for drawing our attention on this aspect.

Figure 1 shows he result of the ATLAS analysis using a cut based analysis, **CBA**, assuming gluon fusion. It shows a significant excess, > 3 s.d., around 650 GeV. Figure 2 shows as well the presence of an excess in the VBF channel using the same CBA analysis. In a so-called **MVA** analysis combining the full information of the events and tuned against the Drell Yan background DY, this evidence seems to disappear at the same rate as the background, while this should not be the case for a scalar resonance.

While ATLAS [8] acknowledges the CBA excesses, they claim that the **ZZ→ℓ⁺ℓ⁻νν̄ search** discards the H650 GeV resonance interpretation. One can however object that the **ETmiss>120 GeV** cut seriously affects a forward peaked ZZ distribution, expected for a spin 2 resonance coming from VBF, as explained in section II.

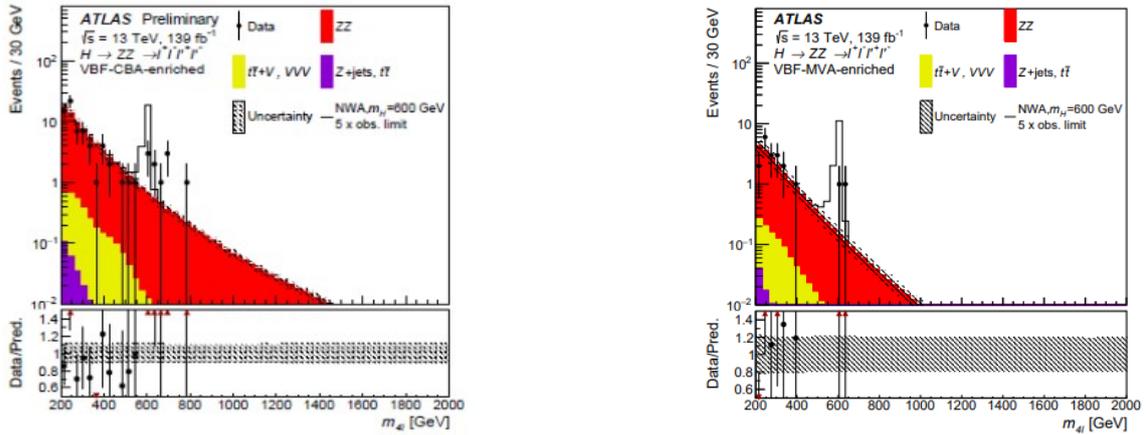

*Figure 2: On the left, the VBF ATLAS (appendix of [9]) CBA analysis applied to the channel ZZ→4ℓ indicates an excess of 10 events around 650 GeV. The MVA analysis [9], on the right, decreases in the same way the excess and the background, while a scalar reference signal at 600 GeV shows that this should not be the case.*

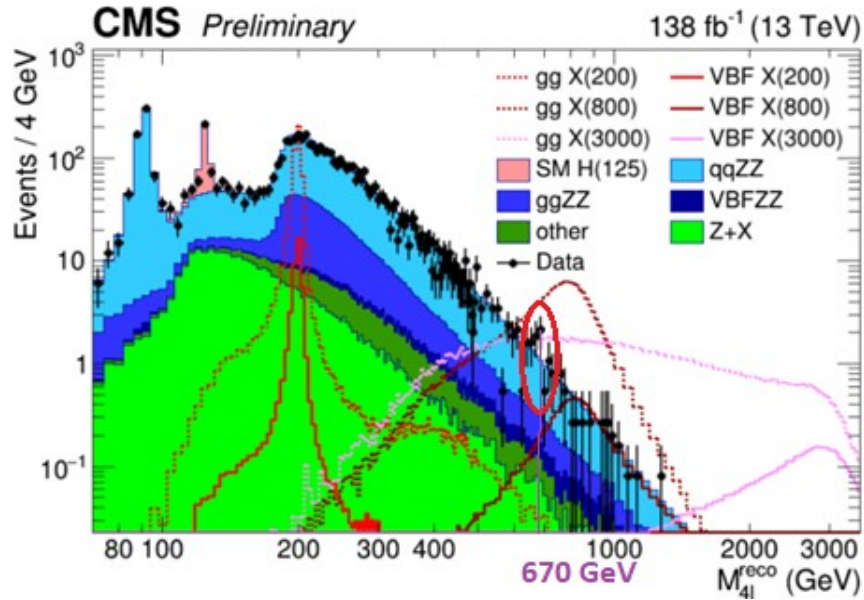

*Figure 3 shows the latest CMS result [10] presented at ICHEP2024 with selections optimised for a scalar resonance, leading to the same conclusion as for the ATLAS MVA result. The red circle shows a mild indication similar to what is observed in the ATLAS MVA analysis.*

## I.2 The W⁺W⁻ channel

Similar inconsistencies occur in this channel. While CMS [11] observes a indication for W⁺W⁻→ℓ⁺νℓ⁻ν, this is not the case for ATLAS. We see two plausible reasons for this. ATLAS only considers μe combinations, decreasing the acceptance by 2 and, more importantly, performs an upper cut of 1.8 on the absolute **difference in pseudo-rapidity** between the two charged leptons, not optimal for a forward peaked distribution of the $W^+W^-$ channel, where the two leptons tend to differ in rapidity.

## I.3 $W^+W^+$ and $ZW^+$ channel

For these channels [12], as we shall see, the VBF angular distribution is not forward peaked, explaining that they are not affected by selections for scalar decays.

## I.4 Possible interpretations

How can one interpret such findings ? On the basis of the sole channel ZZ, one would simply discard the excess as a statistical fluctuation. This natural interpretation however ignores the presence of four other indications observed at the same mass: into $W^+W^-$ [11] by CMS, into h95h125 [13] by CMS, into A450Z [3] by ATLAS[6] and into b$\bar{b}$b$\bar{b}$ [1] by CMS.

In the next section, we will discuss an alternate interpretation which assumes that X650 is a spin 2 particle and that selection methods tuned for the scalar hypothesis H650→ZZ are inadequate. Heavy tensors are predicted within extra dimensions [14] as Kaluza Klein (KK) recurrences of the graviton. In some models, however, one also predicts vector KK recurrences for which LHC has set excellent lower mass limits.

This interpretation may sound far fetched given the present paradigm of searches for extra resonances within the **MSSM model,** as implied by the SUSY scenario so successful in predicting the mass of h125, but recalling that X650 properties contradict such a scenario. In simple terms, within MSSM the **heavier scalar decouples from WW and ZZ,** in clear **contradiction** with LHC observations where X650 couples as a SM scalar to WW. In our previous work [15], we interpret the simultaneous occurrence of $X^{++}$450→ $W^+W^+$ within the **Georgi Machacek** scheme. H650 requires an extension of this model which predicts ZZ/WW=2 for the decay of such a heavy scalar, incompatible with observations. In contrast, a **tensor interpretation** predicts ZZ/WW=0.5 and fulfils the required features for the **WW, ZZ and h95h125** channels. This may not be the case for **A450Z**, which suggests the presence of an **additional scalar resonance at the same mass**.

# II. A tensor interpretation of VBF→T650→ZZ/$W^+W^-$

Firstly, such a tensor can decay into WW and ZZ with a ratio WW/ZZ=2, compatible with observations.

Secondly, in contrast to a scalar, a tensor produced by VBF will decay into ZZ with a forward peaked angular distribution : **d$\sigma$/dcos$\theta_Z$~(3cos²$\theta_Z$-1)²**

This behaviour is similar to the **DY background**, also forward peaked and therefore will not pass efficiently the selections imposed in sophisticated selections against DY. For instance, |cos$\theta_Z$|<0.8 keeps 80% of the acceptance for a scalar but only 46% for a tensor.

This angular distribution is valid for the VBF case. For ggF one has [14]: **d$\sigma$/dcos$\theta_Z$~sin⁴$\theta_Z$** , which should pass the selections for a scalar and one can interpret the absence of signal in the ggF analysis as due to **VBF dominance**. This goes along the interpretation of CMS for the $W^+W^-$ analysis [10] and cannot be excluded from ATLAS ZZ data given the large errors, as explained in Appendix II.

---

6    ATLAS interprets this indication as A650→H450Z on the basis of MSSM.

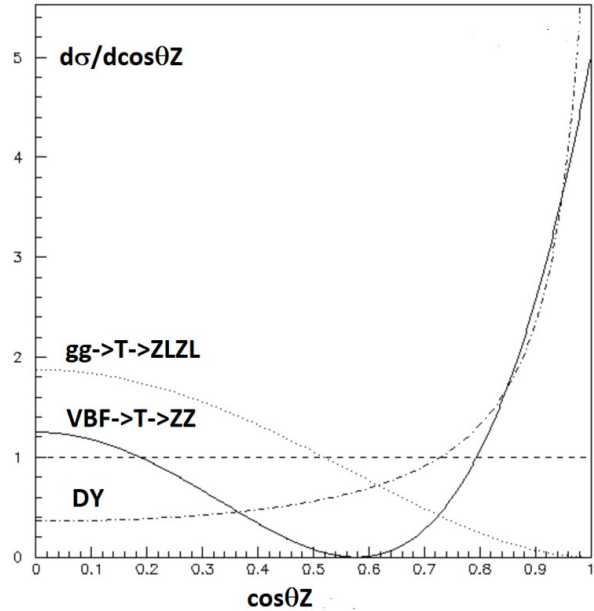

*Figure 4: Angular distributions for a tensor T decaying into ZZ (full curve). The flat dashed curve corresponds to a scalar. The tensor and scalar distributions are given with the same normalisation. The Drell Yan curve DY, dotted-dashed, closely follows the expected distribution for the VBF case.*

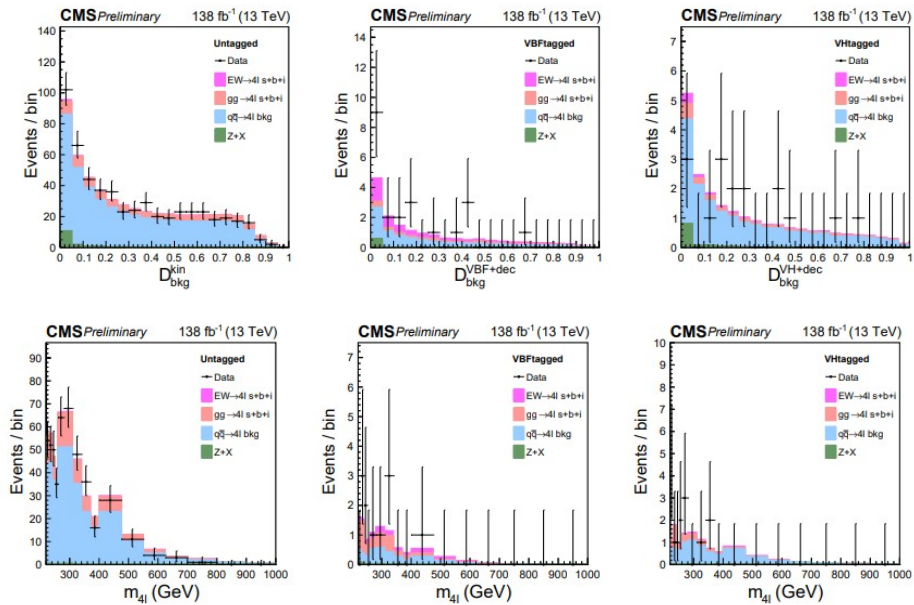

*Figure 5, first row: distribution of the discriminant variable for untagged, VBF tagged and VH tagged events. To enhance a scalar signal over background contributions, a discriminant selection, D>0.6, has been applied to the three categories of the first row, meaning that these distributions only contain a reduced fraction of the total sample, therefore with reduced statistical significance but enriched scalar content. Second row: m4ℓ mass distributions observed after this selection, which show no excess at 650 GeV in the three categories.*

With these arguments in mind, one predicts that sophisticated selections will, as observed, similarly decrease the signal and the background for the VBF part, therefore decreasing the statistical significance, hence the absence of evidence for such a signal in ATLAS with an MVA analysis and in CMS with a similar selection method. In a sense, the 'disappearance' of X650 in the CMS analysis tends to **confirm a tensor interpretation** of this resonance.

CMS has produced an analysis for the reaction ZZ into 4 leptons for an interpretation in terms of h125. To improve on the signal/background ratio, this analysis has used discriminating observables against the DY background [16], with the result shown in figure 5. These distributions show no excess around 650 GeV, from which one would also conclude that the excess measured in that region does not come from a scalar boson. This is only indicative given that the discriminant cut only keeps a fraction of the selected samples but tends to comfort our belief that the discriminating selection applied in [8,9,10] could cause damages to the indication around 650 GeV if it originates from a tensor particle.

For a tensor produced through VBF and decaying into $ZW^{+'}$ one expects: $\quad$ **$d\sigma/dcos\theta_z \sim sin^2\theta_z \, cos^2\theta_z$** which is not forward peaked. For the $W^+W^+$ mode one expects: $\qquad$ **$d\sigma/dcos\theta_w \sim sin^4\theta_w$**

# III. Consequences for the Georgi Machacek interpretation

A tensor resonance does not belong to the GM sector and our attempts to integrate this resonance into an extended version of GM called e-GM appear unjustified. In particular, our analysis in terms of the matrix method is illegitimate since it mixes 3 scalars and one tensor [15].

The **GHW unitarity sum rule for $W^+W^- \rightarrow W^+W^-$** [6] remains valid, where a tensor $T^{++}$ or a scalar $H^{++}$ has to contribute to balance the X650 contribution to the $W^+W^-$ s-channel contribution, meaning that the successful prediction of a **$T^{++}/H^{++} \rightarrow W^+W^+$** contribution keeps its legitimacy in both scenarios.

Given that **$BR(H^+ \rightarrow c\bar{b})$** goes like $Vcb^2$, where $Vcb \sim 0.04$, the **GM model, which corresponds to a type I Yukawa coupling, cannot generate a dominant decay of $H^+130$ into $c\bar{b}$**, as is observed by ATLAS [7]. On the other hand, this light $H^+$ is needed to explain the low BR of $H^{++}$ into $W^+W^+$ by a dominance of $H^+H^+$. A possible way out is to assume that this indication is a fake and that an other light $H^+$ candidate will be found which decays into $c\bar{s}$ and $\tau^+\nu$.

On top of a tension for the decay mode, $H^+130 \rightarrow \bar{b}c$, the GM solution encounters an apparent contradiction with precision measurements:

- The value of the expected vacuum energy expectation of the two GM isodoublets deduced from the GHW sum rule involving $W^+W^+$ and $ZW^+$ is above 70 GeV, which seems incompatible with the reported value of 30 GeV of [18]

- A similar contradiction is observed with the predictions from b physics [19] which are also in tension with this high value of this vacuum expectation

- The triplet mass m3 of order 140 GeV seems inconsistent with the interpretation of direct searches from LHC of [18].

# IV. A tensor+scalar solution

The various tensions and inconsistencies described in the previous section lead us to **abandon GM** and assume that $T^{++}450$, $T^+375$ and $T350$ form an **isoquintuplet of tensors**. We will assume that X650 is presumably also part of an isoquintuplet, yet incomplete, and collect the remaining candidates into **isodoublets of scalars**. Three of them will be needed, leading naturally to a **3HD model** (see Appendix I)**.**

## IV.1 Inputs for a tensor scenario

The present section is meant to describe some features relevant for a tensor (spin 2) scenario.

First of all, in a tensor scenario one should take into account the presence of the **spin multiplicative factor 2J+1=5** which enters in the calculation of the cross section:

$$\sigma_f \sim (2J+1)\Gamma_i BR_f$$

where, in the case of VBF, $\Gamma_i = \Gamma_{WW/ZZ}$ and $BR_f$ is the branching ratio of the resonance in the final state.

From the VBF cross section for WW ~160 fb (Appendix II), one infers $\Gamma_{WW}$=15 GeV, instead of 70 GeV for the scalar case. It seems therefore that **interference effects** could be faking an apparent total width of 100 GeV. We will come back to this issue in section IV.4.

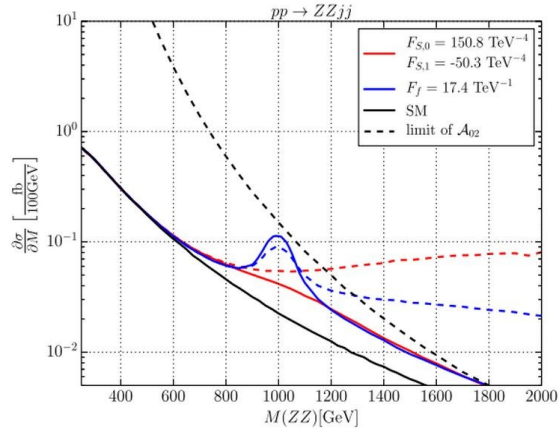

*Figure 6: Differential cross section for an isoscalar spin-2 resonance with mass 1000 GeV and width 100 GeV, for VBF→ZZ. Solid line in blue : unitarized results, dashed line uncorrected results.*

The **U.V. behaviour** of tensor resonances is a notorious issue. It can be taken care of by the so-called **'unitarization procedure'**. From figure 6 [20], one retains that the unitarized solution, in blue, behaves as a genuine BW, without blowing up at high mass. This means that the tensor contributions to GHW sum rule for $W^+W^- \to W^+W^-$ remain meaningful with the important prediction of resonances of type $T^{++}$.

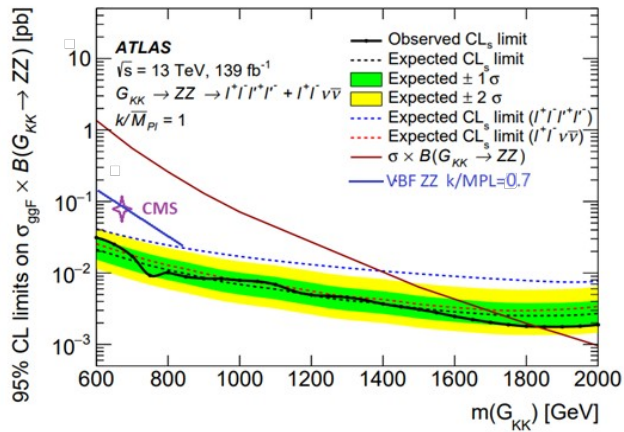

*Figure 7: From ATLAS [8], expected and observed 95% CL upper limit on resonant production cross section for spin-2 pp→X→ZZ signal hypothesis. The red line shows the bulk KK graviton predictions for $\kappa/\overline{MPl}$=1. The blue line shows the prediction for the VBF cross sections for $\kappa/\overline{MPl}$=0.7 (see Appendix II). The magenta star shows the CMS result deduced from VBF→X650→W⁺W⁻. Recall that the upper limits from ATLAS assume an acceptance invalid for the VBF case.*

Keeping only the resonant Born term, one has the following amplitude:

$$A(w+w- \to w+w-) = -(1/24)F_t{}^2\, s^2 P_2(s,t,u)/(s^2-m^2+im\Gamma_{tot}) \text{ with } P_2=[3(t^2+u^2)-2s^2]/s^2$$

which, in the relativistic limit, confirms the VBF angular dependence given in section II.

For the tensor case, one has ZZ/WW=0.5, not incompatible with experimental observations given the large uncertainties in the determination of the VBF and ggF cross sections (see Appendix II).

## IV.2 A Randall Sundrum scenario

Leaving generalities, one may ask if such a tensor solution could fit into some specific BSM models. What comes to mind are **Randall Sundrum models** [14] with **warped extra imensions** which predict heavy Kaluza Klein excitations of the **graviton** $G_{KK}$.

The experimental situation is summarized by figure 7 [8] where one sees that the RS prediction, usually assumed dominated by the ggF contribution, excludes a KK graviton lighter than 1.8 TeV. This problem is alleviated if one assumes that this resonance is produced mainly via VBF, as suggested by the CMS analysis for the $W^+W^-$ channel.

RS predictions for $G_{KK}$ couplings to SM particles are widespread in the literature [21,22,23,25]. $G_{KK}$ is assumed to be localized near the **TeV brane**, as well as the top quark, the Higgs boson and, through the equivalence theorem, the $Z_L/W_L$ components of these bosons. Therefore one expects this particle to decay primarily into these particles and be produced by VBF through **$W_L W_L$ and $Z_L Z_L$ fusion,** as suggested by the CMS analysis.

One also expects an important **ggF contribution** originating from two sources:

- The standard **top loop contribution** which is model dependent. This problem is solved in [22] where **top quarks are "sequestered" from KK graviton** in the extra-dimension

- The **tree-level ggF contribution**:

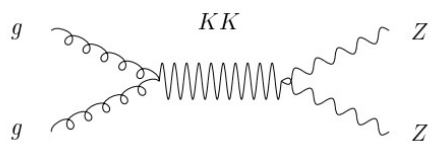

which seems to give a ggF contribution **in excess with observation**. One therefore tends to conclude that data look inconsistent with standard RS models. As discussed in section IV.4, one can assume a **non universal coupling of the KK particles** which would favour ZZ/WW coupling over gg and, presumably, $\gamma\gamma$ couplings, the later with reduced BR, explaining why KK has escaped detection in this gold plated mode.

Within RS, a scalar **radion r** is expected to couple to $G_{KK}$ as a longitudinal boson ZL. One can speculate [26] that h95 and h125 carry a radion component, explaining the observation of X650→h95h125 by CMS.

The $G_{KK}$ resonances have **quantified masses** [22] which go like $mi \sim x^G{}_i$ where: $x^G{}_i$=3.83,7.02,10.17,13.32. The isotensor T350 would be the first KK recurrence, T690 the second one (this modified mass follows from the discussion in IV.3). T1000, the third and T1300 the fourth are still unobserved. Under this hypothesis T350→ZZ was not detectable since its coupling$^2$ is $(7.02/3.83)^2$=3.35 times weaker, with larger SM background. Presumably the BR into ZZ remains the same. The SM background is 5 times larger as shown in figure 1. It is likely that, with increased integrated luminosity, one could detect T350→ ZZ directly or in the process A450→T350Z.

Note that the width of T690 could be of order 20 GeV (see IV.3), meaning that recurrences T1000 and T1300 could have widths of 40 GeV and 70 GeV, therefore be observable with more luminosity. Note also that these resonances will not only be affected by **interferences** with the SM backgrounds as seems the case for T690, but between themselves, since they share the same final states and, presumably, are produced by VBF.

Again, this hypothesis has also consequences concerning the GHW sum rule since this rule relates square couplings to ZZ which are growing functions of the rank of the KK resonance. This effect may however be absent due to **isospin symmetry** as discussed in Appendix III.

Appendix II gives an evaluation of **BR(T$^{++}$→W$^+$W$^+$)~0.2** using the predictions of the couplings given by the RS model. This calls for the presence of a light H$^+$ to provide a large BR(T$^{++}$→H+H+).

### IV.3 A narrow tensor resonance T690 observed into two photons?

A heavy SM Higgs boson is a wide resonance dominated by WW/ZZ/t$\bar{\text{t}}$ modes, which only receives minute, therefore unmeasurable, loop contributions, from the two photon modes. This is not the case for a KK graviton tensor which couples directly to the energy-momentum tensor containing photon, gluon, Z/H/W, fermions contributions [32] (see Appendix III for an explicit expression of this tensor):

$$\mathcal{L}^{G}_{V,f} = \frac{k_{V,f}}{\Lambda} \, T^{V,f}_{\mu\nu} \, G^{\mu\nu}$$

From LHC observations, one expects the following VBF production mechanisms:

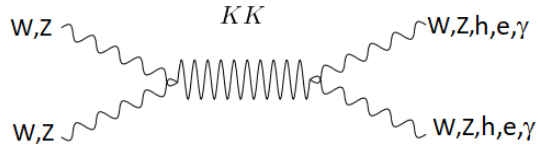

with an angular distribution ~ (3cos$\theta^2$-1)$^2$ for W,Z,H, and ~ **sinθ$^4$** for photons, the later being eminently favourable for detection.

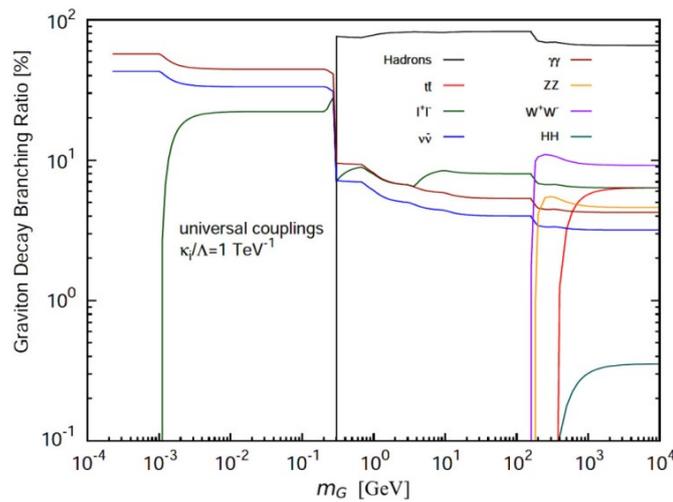

*Figure 8: Graviton decay branching ratios [32] assuming a universal coupling of gravitons to the energy-momentum tensor.*

Assuming a universal coupling of the graviton to the various SM channels, one predicts the BR given in figure 8.

ZZ/WW and γγ decays can therefore interfere with various SM and BSM diagrams:

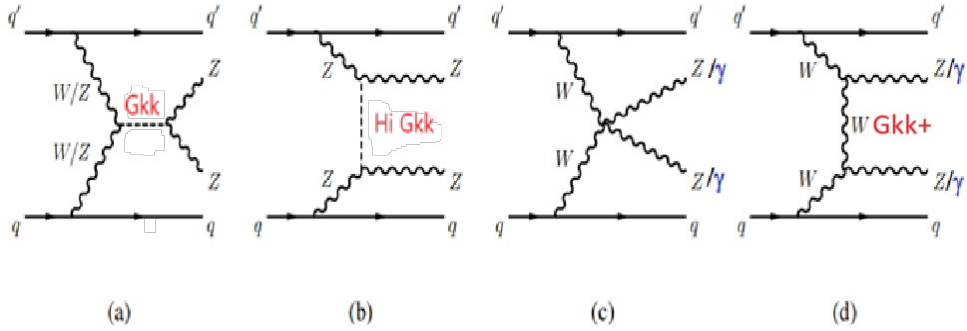

where $G_{KK}$ is the sequence of KK gravitons among which T350, T1000, T1300, ... and Hi are the CP-even neutral scalars of the 3HD classification discussed in the next section. On top of these t and s channel exchanges, there are u-channel tensor exchanges, with double charge for WW→WW and single charge tensor for WW→ ZZ. For the two photon final state process (d), can proceed with a charged KK graviton.

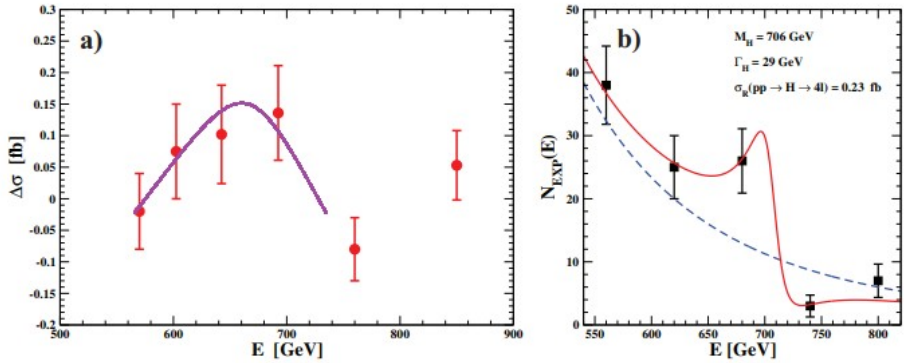

*Fig 9: Plot a) from [34] shows an analysis of the ZZ data from ATLAS interpreted without taking into account interference effects, which suggest a wide resonance centred a 650 GeV while plot b) shows a narrow resonance centred at 700 GeV.*

Below the T690 resonance, its real part is negative and interferes destructively with the s-channel amplitude of T350 which is positive therefore shifting the apparent resonant mass to a higher mass value. The opposite is true for T1000, T1300 etc... ,the t-channel (b) and (d) and the u-channel exchanges which are dominant. Figure 9 confirms this with an empirical fit of the interfering background.

With ATLAS data, [34] finds a mass of 700 GeV and a width of 29 GeV instead of 650 GeV and 100 GeV suggested by figure 9-a. **CMS data** lead to a **total width of 16 GeV** and a mass of 690 GeV [34].

Again we reiterate our plea in favour of a CBA analysis of these data in order to be compatible with a spin 2 resonance. When this result will be available and with a better knowledge of the couplings to ZZ of h95 and H650, one should be able to perform a fit taking into account interferences.

Note that (c) and (d) will interfere with **WW→T690→ γγ.** Figure 10 shows a **narrow signal in γγ,** indicating that interference effects are small for this channel which therefore delivers the **best estimate of the mass of this resonance,** consistent with the fits of [34]**.** The smallness of the interference effect can be understood from the angular distribution of this resonance which goes like **$\sin^4\theta$**, which therefore has little overlap with the forward peaked distribution of diagram (d).

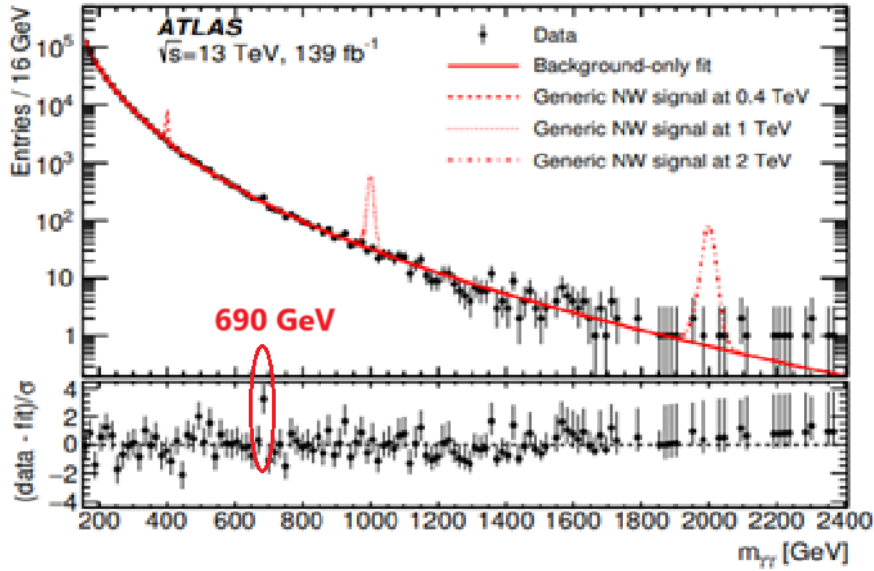

*Figure 10: ATLAS search for narrow resonances decaying into γγ [33]. One observes a 3.5 s.d. deviation in the 690 GeV bin which corresponds to an excess of 50 events, that is s/b=1/4.*

<u>Taking into account these various features, from now on we will label this resonance **T690**, which says that it is a tensor with a precisely known mass of **690±10 GeV** and a total width of order 20 GeV.</u>

These conclusions are based on the 4 lepton and the two photon channels with **excellent mass resolution**. Selecting ZZ with neutrinos or jets, with **poor mass resolution,** seems therefore inadequate.

The BR deduced from figure 10 is **two orders of magnitude below the measured value for ZZ,** contrary to the universal coupling picture assumed in figure 8. This behaviour roughly agrees with [35] where BR(γγ) is predicted 20 times smaller than the BR(ZZ) (see Table 1 in [35]). Quoting K. Agashe: 'For a zero-mode graviton, couplings would be universal, but that principle does not apply to a KK graviton'.

Table 2 of [35] also predicts a **Z_KK gauge boson with a mass of 3 TeV,** a cross section of 3fb, a BR into di-bosons (H/W/Z) between 0 and 44% and BR(e+e-) between 0 and 3%. Reference [36] shows some evidence for di-boson resonances around 2.1 and 2.9 TeV with a few fb cross section times BR, at the 3.6 s.d. level.

For e+e-, [35] predicts up to 6 Zkk→e+e- events at RUN2, therefore observable at LHC with more luminosity. One could also look for **W_KK,** noting that CMS has an indication for **W_KK→ tb̄** at a mass of 3.8 GeV [37].

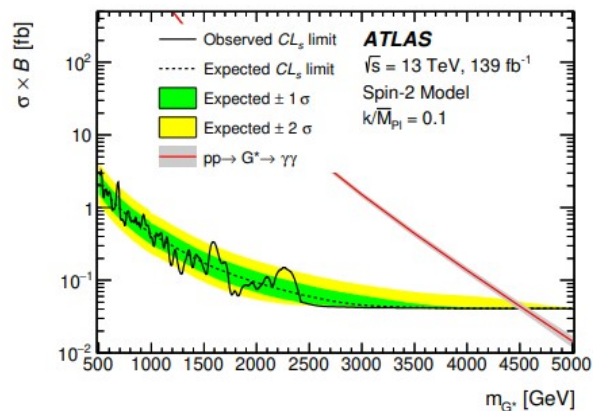

*Figure 11: Upper limits from ATLAS for spin-2 resonances cross sections times branching ratio into γγ [33].*

If the coupling to two gluons also departs from universality, one can understand why **VBF is dominating over ggF for the production of the KK graviton**.

The absence of signal for **ℓ+ℓ-** implies that this could also be true for leptons. For a universal coupling, one predicts a Br(e+e-) equal half BR(γγ). We will come back to this issue in section V.

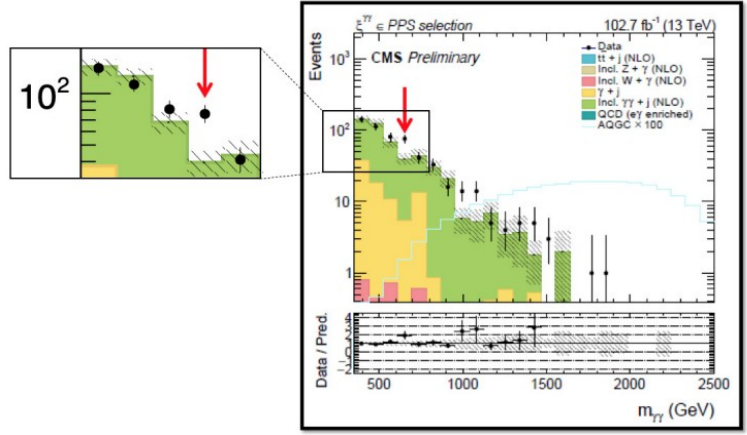

*Figure 12: From [34], number of photon pairs reconstructed by CMS in coincidence with TOTEM [37]. The 650±40 GeV bin shows an excess of 36 events, that is s/b=1, with a significance of 3.5 s.d. The bin width is 80 GeV.*

**CMS and TOTEM** detectors [38] are combined to tag a two photon initial state producing a heavy KK graviton with subsequent decay into two photons, to be detected by the CMS detector. The resulting spectrum is shown below and, as noted in [34], it shows a 3 s.d. excess at 650±40 GeV as one would expect from the decay of T690. In practice, most of these events are random coincidences and, as expected, only one event shows an agreement between the masses reconstructed in the two detectors.[7]

A fully quantitative interpretation of the CMS-TOTEM effect goes beyond the scope of the present paper. If one compares the excess of γγ events reported by A̅T̅L̅A̅S̅ to the result shown in figure 12, it seems that γγ events produced by VBF have a large probability that the **accompanying forward jets trigger TOTEM**.

## IV.4 An extra X650 candidate and its interpretation

Recently an indication [1] has been reported by CMS for **X700→Y400h125→ bb̄bb̄.** Given the poor mass resolution, it can also originate from **T690→T350h125 or from H650→T350h125.** The later seems more likely since the total width of T690 seems already saturated by the ZZ/WW and hh modes.

The decay of T350 into bb̄ may come as a surprise since we have assumed that T690 is decoupled from tt̄. In the genuine RS model, one expects that the preferred decays of $G_{KK}$ are through top and b quark pairs, dominating over hh and ZZ/WW modes. For T690 however, the ggF process seems to be subdominant, implying that top quarks are decoupled due their distribution in the extra dimension [22]. This may not be the case for b quarks, explaining the large coupling of T350 to b quarks, as implied by the new candidate.

---

7    Many thanks to C. Royon for this information.

# V. Proposed global interpretation

## V.1 The 3HD sector for scalars

Having rejected our previous attempts within GM or e-GM, there remains to propose an alternative classification of the genuine scalars indicated by LHC data. We adopt the three Higgs isodoublets (3HD) model already introduced a long time ago by Weinberg [28] and reconsidered recently [27] to interpret an indication for a light $H^+$ in ATLAS.

This 3HD model can accommodate the various **scalars** suggested by LHC findings: **h125, h95, $H^+$130, A152, H650, A450** and an additional heavy **$H^+$700,** interpreted as A450W$^+$ [17], whose mass is predicted above 700 GeV [27]. One also assumes that there is a **neutral scalar resonances around 650 GeV**, observed in AZ and in 4b, . This hypothesis allows to understand why these channels do not contribute to the **total width of T690** which is nearly saturated by ZZ and WW contributions.

The 3HD model also provides, in a restricted parameter space [27], an interpretation for $H^+$**130→$\overline{b}$c**, satisfying the various constraints from searches and from precision measurements, primarily b→ s γ. It predicts that BR(t→H$^+$b)BR(H$^+$→c$\overline{b}$)=0.14%, as explained in Appendix I. The **second charged Higgs** should be heavier than 700 GeV, compatible with the mass of 650 GeV suggested for the neutral scalar of the heaviest isodoublet, as indicated in the table below. In the Kaluza Klein interpretation, one predicts (see Appendix II) that **BR(T$^{++}$450 ->W$^+$W$^+$)=20%**, hence the need for a **light $H^+$** to provide H$^+$W$^+$ and H$^+$H$^+$ decay modes for T$^{++}$450.

## V.2 The RS sector for tensors

This model predicts heavy KK recurrences of ordinary SM particles. To satisfy the SM precision measurements, in particular the **ρ parameter constraint**, one usually concludes that these particles are very heavy. Additional symmetries allow to alleviate this constraint and allow KK recurrences of Z/W down to 3 TeV, therefore reachable by HL-LHC.

Coming back to KK gravitons, in [35] it was shown that, assuming **two intermediate branes,** it is possible to predict **G$_{KK}$ gravitons lighter than all other KK particles**. LHC observations suggest that the two first KK excitations would fall, respectively, at 400±50 GeV and 690±10 GeV. One would therefore predict the next excitation at 1 TeV. For the first excitation, one observes two charged partners, T++ at 450 GeV and T+ at 370 GeV. This first generation would therefore form an **isoquintuplet.**

For the 2$^{nd}$ excitation one only observes a neutral candidate but it is tempting to believe that the charged candidates have been missed due to more complex topologies which open up at 700 GeV. An other motivation for this hypothesis is that it would allow **perfect cancellations in the GHW unitary sum rule** through the u-channel exchanges allowed by the charged tensors (see Appendix III).

Indeed, the issue of satisfying **unitarity** has already received some attention in particular for the graviton sector [36]. If each KK excitation would form such multiplets, one would elegantly solve this problem.

This is therefore a critical prediction of the present work: **one expects to observe the 3 first KK excitations of the graviton in neutral, charged and doubly charged modes.**

## V.3 Summary

The GM model is ignored and one is left with the **3HD model** and two **tensor isoquintuplets TQ1 and TQ2**. Note in passing that this 3HD solution would be discarded in [29,30,31] since it cannot fulfil the unitarity criteria given that [30] ignores the contribution of the KK part (see however Appendix III).
This scenario has several merits :

- It integrates all BSM candidates
- It solves the T690→ZZ puzzle
- It includes $T^{++}{\to}W^+W^+$ and $T^+{\to}W^+Z$ predicted by the sum rule and indicated by LHC
- It predicts $T^{++}{\to}H^+130H^+130 \to \overline{b}c\overline{b}c$
- It predicts a narrow T690 resonance
- It predicts T690→γγ
- It predicts charged KK graviton recurrences at 700 GeV
- It predicts the following graviton isoquintuplet TQ3 at 1 TeV
- It preserves unitarity in the KK graviton sector

The table below summarizes our interpretation.

| HD1 | $h125{\to}SM\ modes$ | **WL** | **ZL** |
|---|---|---|---|
| HD2 | $h95{\to}2\gamma$ | $t{\to}H^+130b{\to}\overline{b}cb$ | $A152{\to}2\gamma + Z/W\ from\ T^+/T$ |
| HD3 | $H650{\to}t\overline{t}Z$ /$T350h125$ New | $H^+700{\to}A450W^+{\to}t\overline{t}W^+$ | $A450{\to}T*350Z{\to}hhZ^8$ |
| TQ1 | $T^{++}450{\to}W^+W^+$ $\to H^+H^+$ ? | $T*375{\to}ZW^+$ | $T350{\to}hh,b\overline{b}$ |
| TQ2 | $T690{\to}ZZ/WW/h125h95/\gamma\gamma$ **New** | | |

Concerning **A152**, so far observed[9] into 2 photons accompanied by b quarks, leptons or missing transverse energy, it can be interpreted as a cascade originating from the tensors. One should therefore select these topologies and identify them as coming from a cascade involving Z bosons. This should allow to reconstruct the parent(s).

A650→H450Z→tt̄Z from ATLAS, which can also be understood as H650→ A450Z, shows a large cross section interpreted as coming from the process $b\overline{b}{\to}A/H$. Quantitatively ATLAS measures a CL upper limit of 636 fb for an expectation of 257 fb. Such a large excess seems in tension with the $t\overline{t}Z$ measurements from ATLAS and CMS which agree with the SM.

---

8    A450 is a resonance observed in top pairs which allows proving its CP-odd nature, which therefore eliminates a graviton interpretation.

9    See arXiv:2503.16245 for a very recent update.

# VI. A graviton factory scenario

## VI.1 Arguments in favour of a KK graviton

To summarize:

- LHC observes that T690→ZZ is inconsistent with J=0, suggesting a J=2 interpretation of this resonance

- At LHC, the production mechanism goes through VBF fusion through $W_L W_L / Z_L Z_L$ coupling. Decay modes, through ZZ/WW, HH and $\gamma\gamma$, also agree with predictions

- This resonance is narrow, with a total width of 20 GeV which can be quantitatively interpreted within RS

- In a KK interpretation, one predict a sequence of states with predicted mass ratios. It one assumes that T690 is the second KK resonance, one expects the third resonance at 1 TeV and the first one at 400 GeV. The later has been observed in the form of an isomultiplet T350, T+375 and T++450

The remaining open issue is the coupling of a graviton to e+e- which, as we shall see, can have dramatic experimental consequences. If one assumes that T690 is a KK recurrence of a graviton, figure 8 clearly suggests that this resonance will directly couple to e+e- with a coupling constant similar to the two photon coupling.

It is however true that this picture contradicts the **bulk description** of the RS model where one assumes that light fermions reside on the Planck brane and therefore do not couple to KK gravitons which peak at the TeV brane. This was not the case with the primitive version of RS where all SM particles were residing on the TeV brane, while only the KK gravitons were propagating in the bulk with a peaked distribution near the TeV brane. This first vision has justified the initial search for KK gravitons into the e+e- mode.

Taking an **agnostic attitude**, we will examine available LHC data to see if there is any sign of a signal around 690 GeV in e+e-.

## VI.2 Observing T690 -> e+e-

While, for heavy resonances, LHC detectors have an excellent mass resolution in **e+e-**, similar to $\gamma\gamma$, this is not the case in **μ+μ-**. We should therefore concentrate on e+e- since T690 is a **narrow resonance** affected by interference effects.
The level of e+e- background at a 690 GeV mass is about twice larger than for $\gamma\gamma$, which a priori allows a similar level of sensitivity. A major difference concerns the expected **interference effects**. While apparently the two photon channel shows almost no effect, the e+e- channel could behave differently, which can explain why it has been unnoticed so far.
Figure 13 and 14 from ATLAS definitely show an observable effect at the right mass. A sophisticated analysis taking into account interference effects is needed to quantify the cross section. This clearly belongs to the LHC experiments. Crudely speaking this cross section is similar to the two photon cross section.

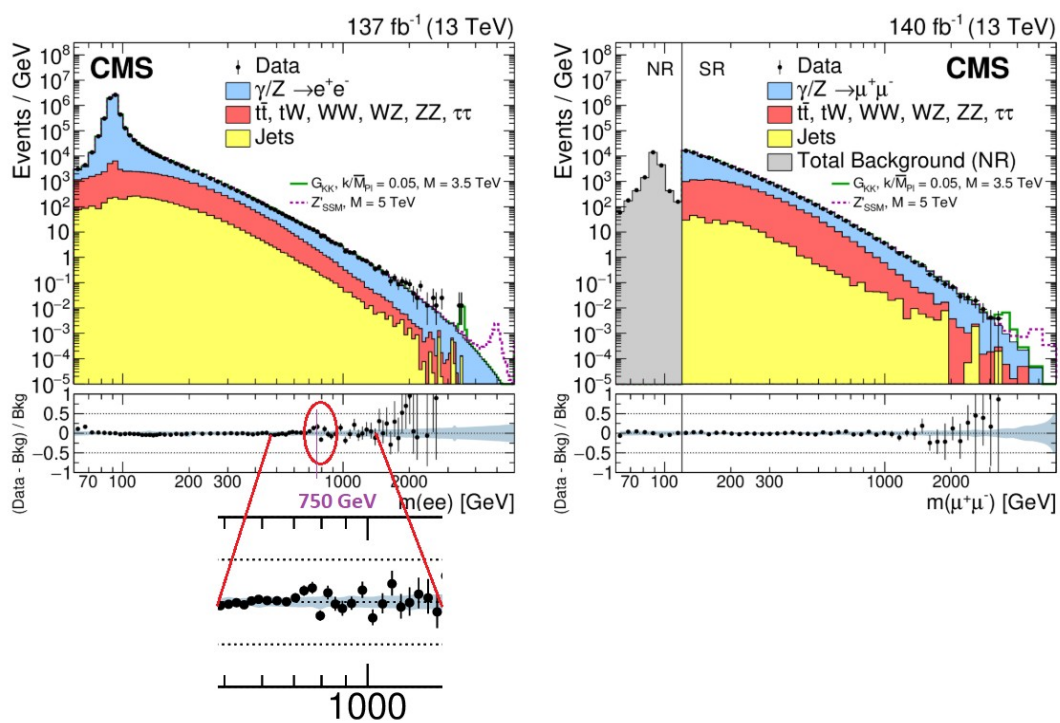

*Figure 13: e+e- and μ+μ- mass spectra from CMS [39].*

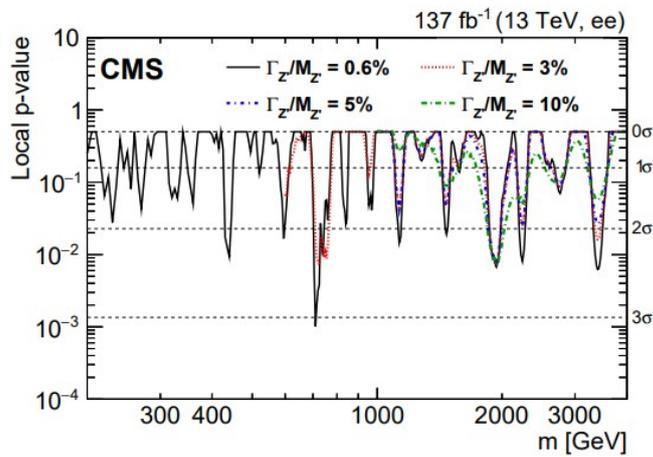

*Figure 14: Observed p-value for the channel e+e- with four values of the mass resolution.*

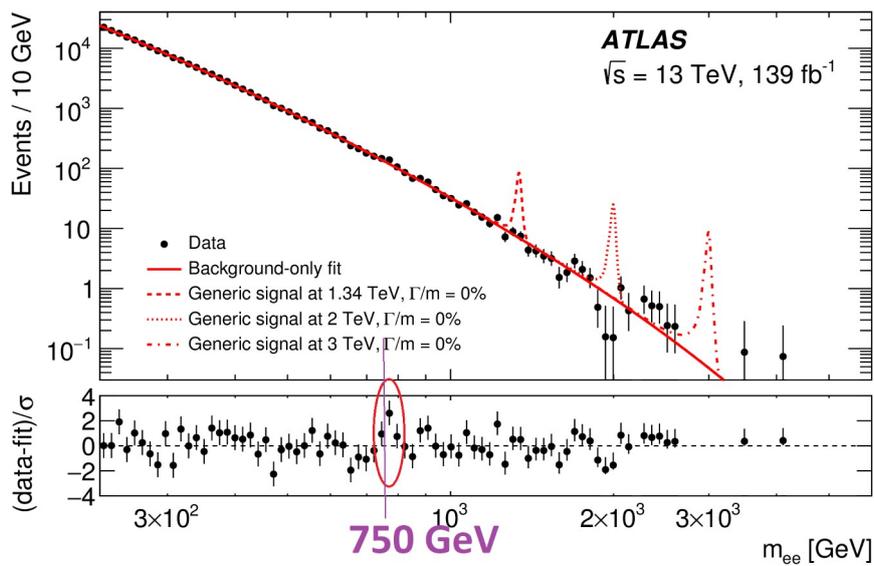

*15: e+e- mass spectrum from ATLAS [40].*

ATLAS data indicate an effect at the same mass, as shown in figure 15 from [40].

One therefore tends to conclude that the KK graviton candidate T690 does couple to e+e-. Keeping the RS bulk description for the interpretation of fermion masses, one may suggest that since the mass of an electron results from the overlap, in the extra dimension, of the Higgs distribution peaked at the TeV brane with the distributions of the e-L and e-R components (see for instance [42]), one can generate this low mass by tuning the eL distribution peaked towards the Planck brane while the **e-R component** could have a larger overlap with the Higgs distribution. If this idea is correct, one therefore predicts that in e+e- annihilations, T690 will be predominantly produced by e-R. This interpretation has its limits given the accuracies reached on e-R couplings at the Z pole [43].

We will come back in section VII to what can be expected at an e+e- collider if T690 couples to e+e-.

# VII. Triple Higgs coupling measurements

These measurement are presently the main focus within ATLAS and CMS.  As predicted by [41], the presence of H650 could have some consequences on the value of $\kappa_\lambda$. This reference claims that scalars with intermediate mass scales and with sizeable couplings to the Higgs vacuum expectation will modify the Higgs potential and therefore the value of $\kappa_\lambda$. The heavier they are, the larger this deviation, an unusual blessing which reminds us of the influence of the top quark mass on the $\rho$ parameter. This is illustrated in figure 16 from [38] where the variation is shown for several models.

The main issue here is **model dependence**[10] of this result. For instance, within **MSSM**, the decoupling limit $\sin^2(\alpha-\beta)=1$ insures SM couplings for hWW/ZZ, irrespective of H mass. At **tree level,** one has $\kappa_\lambda=(mz/mh)^2\cos^2 2\alpha$, that is **$\kappa_\lambda$ < 0.5**, independently of the heavy scalar mass. This illustrates that MSSM behaves distinctly from generic 2HD models where couplings vary freely, only limited by unitarity constraints, while in MSSM they are given by coupling constants. This model has therefore the virtue of predicting successfully the lightest Higgs mass mh, up to **loop corrections**, which are large given that, at tree level, one expects mh<mZ. For $\kappa_\lambda$, these loop corrections are related to those of mh and one expects **$\kappa_\lambda$~1**. It is fair to recall that MSSM drastically differs from the 3HD model suggested by LHC data. In our interpretation of H650 belonging to the heaviest HD, one can in principle precisely compute $\kappa_\lambda$.

The $\kappa_\lambda$ parameter measurement can also be affected by **BSM resonances** decaying into hh pairs. This would be the case for T350 observed by ATLAS [28] in the cascade A(450)→T350Z→hhZ. If confirmed, this resonance should be taken into account to measure $\kappa_\lambda$. This can be understood from figure 17 where one can predict that this resonance would fall in the middle of the mass distribution of the SM contribution. Noteworthy, the size of this cross section could be **twice larger** than the SM contribution which is a the level of 30 fb for the ggF part. The CL upper limit for the ggF cross section A→ZH→Zhh→Zb$\overline{b}$b$\overline{b}$ is of order 70 fb, implying a significant contribution of this BSM source which needs to be subtracted to reach the genuine value of $\kappa_\lambda$.

Interestingly, a recent update of the di-Higgs analyses presented at Moriond [45] concludes that, for Run 2, upper limits for **ggF** production are about the same for the two experiments: **2.9(3.5)xSM for ATLAS(CMS)**, in slight excess (2.5xSM expected), therefore still compatible with the presence of the **A→Zhh** channel indicated by ATLAS. Note also that, for **VBF**, the upper limit reached by CMS is of order **100xSM**, meaning almost insensitive to the contributions coming from **KK gravitons** produced through this process.

---

10   Thanks to Gilbert Moultaka for pointing out this aspect

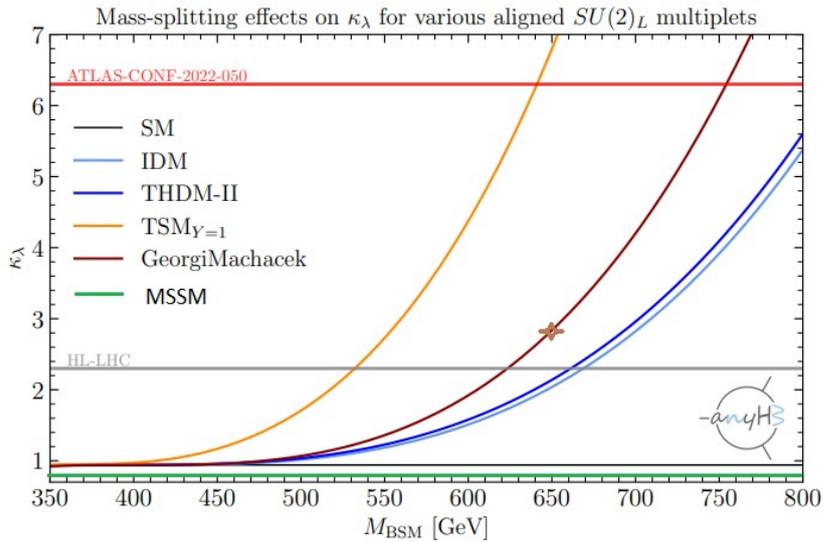

*Figure 16: κλ versus heavier scalar masses for various models [41]. The green line corresponds to a MSSM solution with tanβ > 5. The GM solution, κλ~3, is indicated by a star.*

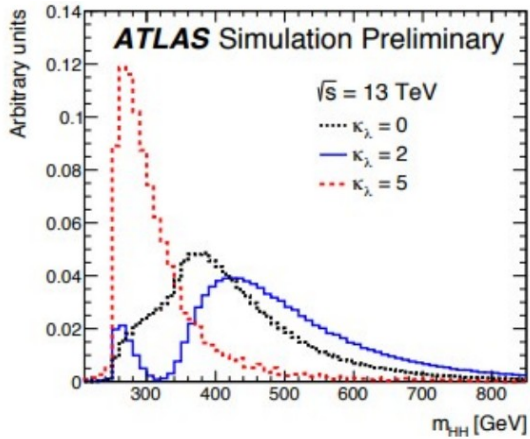

*Figure 17: Mass distributions of the hh final state for various values of the parameter κλ*

# VIII. Prospects at future colliders

If we are right, BSM channels like T690 are rich and complex and deserve the highest attention from the HEP community. A **linear collider** reaching 1 TeV is clearly able to measure the heavy tensors which are of major interest. ILC could deliver 8000 fb-1 at this energy. Typically, processes like T690→ AZ→hhZZ are complex multijet channels, requiring a **high performance detector,** offering excellent particle identification, including b charge measurement, over the **full solid angle**.

Given the poor determination of the ggF and VBF cross sections achieved so far at LHC, it is presently difficult to accurately predict the cross sections. In the 3HD scenario (see Appendix I) one can produce h95 in association with A450. A152 can be observed from cascades or in association with h95.

In section VI we saw that there are indications that **T690 couples directly to e+e-**. If one assume that, similarly to photons, the coupling is ten times lower than for ZZ, one expects BR(e+e-)=0.25%, which gives a **cross section of 150 pb** for a center of mass energy of 690 GeV. This means that a LC collecting 1000 fb-1 at 690 GeV would produce $10^8$ **events !**

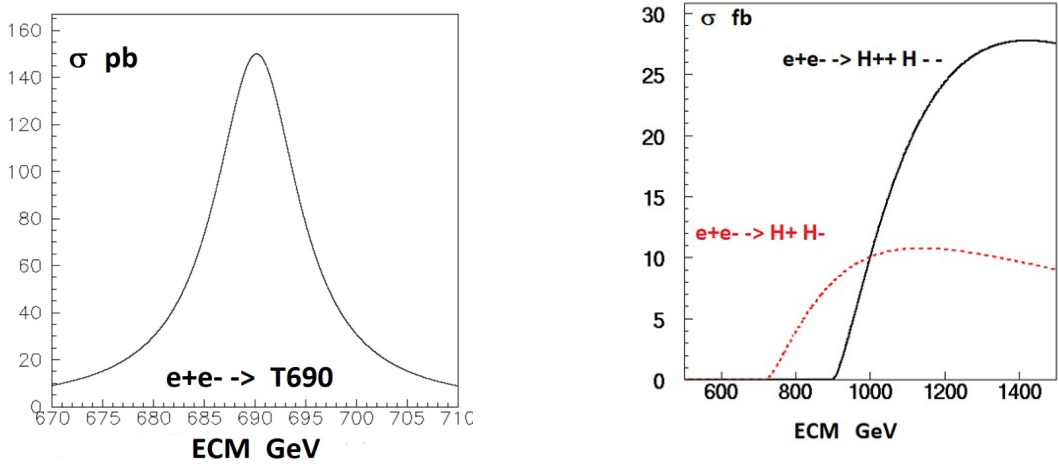

*Figure 18: e+e- cross sections for heavy resonances*

A similar scenario could occur for **T350** although there is no sign of a coupling of this resonance to e+e-, which can be understood by a higher level of background at lower masses. One predicts (Appendix II) that the BR(T350→e+e-) remains the same while, according to the RS model, the total width is reduced by the ratio $(3.83/7.02)^3$ which gives 3 GeV. The e+e-→T350 cross section at resonance could reach **600 pb**. Similarly, **T1000** could be reachable at a TeV LC, providing a third KK graviton factory.

An **e⁻e⁻ facility** would allow to provide a large cross section for T⁻⁻ through **W⁻W⁻ fusion**. A γγ facility would allow to directly produce and measure **γγ →T690,** but is not competitive if T690 couples to e+e-.

The pair production cross section for $e^+e^-\rightarrow T^+T^-$ differs from $e^+e^-\rightarrow H^+H^-$ given in figure 17 by having 25 final state **spin configurations** instead of 1. While both channels have the same coupling **eQ** to the photon, charged tensors couple to Z bosons like **e(I3-Qs²w)/swcw**, were I3=1 for T⁺ and I3=2 for T⁺⁺. These two effects will therefore result in a **large enhancement** with respect to the scalar case and should provide precise measurements[11] in e+e-.

A **circular machine** can access to h95, A152, H⁺130 and T350 (if it couples to e+e-). Identifying H⁺130 in an e+e- collider will be trivial below the top pair threshold through the process e+e-→Z/γ→H⁺H⁻. In the 3HD scenario (see Appendix I), one can produce h95 in association with A152, completing the second isodoublet identification. In this scenario one can also measure e+e-→Z→h95h125, providing a direct proof of CPV in the Higgs sector. Indirect sensitivity to heavy scalars through $\kappa_\lambda$ can also be achieved using the h125Z channel.

**FCChh,** even operating at 50 TeV, would seem an **ideal machine** to further explore the **KK graviton spectroscopy**, reaching **higher rank KK recurrences of the graviton** and, eventually, discovering **KK vectorial recurrences**, with masses circa 3 TeV. This would definitely elucidate a RS interpretation.

---

11  To our knowledge, this cross section is unavailable in the literature.

# IX. Conclusions

**LHC could provide an invaluable exploration of the mass domain which can be covered with a TeV LC.**

So far, T690, the most significant candidate for BSM physics at LHC, has suffered from two misinterpretations:

- It was analysed assuming J=0, which resulted into **inadequate selections,** reducing its significance

- It was interpreted as a **wide resonance**, ignoring interference effects with the SM background

- It seems mass degenerate with the scalar resonance H650

This note is an update of our work on the interpretation of various indications observed by ATLAS and CMS for new resonances which has taken into account these aspects. It was triggered primarily by the claim of CMS of an absence of indication for H650 into ZZ. We provide an interpretation for this absence as due to a **wrong hypothesis** on the spin of this resonance. We claim that T**690→ZZ originates from a J=2 resonance produced by VBF**. In such a scenario, the treatment applied assuming that T690 is a scalar is inadequate since selections for VBF→T690→ZZ have about the same effect for signal and background, also explaining the behaviour observed by ATLAS in their sophisticated ZZ selections for a scalar particle.

**Interference effects** tend to hide a narrow T690 resonance, as confirmed by $\gamma\gamma$ and e+e- channels, which clearly calls for a **Kaluza Klein graviton interpretation.**

This result cannot however be interpreted within the strict **standard RS** framework since:

- There is no sign of the large **ggF contribution** as expected within this scheme

- There is no sign of vector recurrences in the LHC data, as one would expect in such a model

On this last point, [35] predicts that one can have a hierarchy like $M_{KKHiggsmatter} > M_{KKgauge} > M_{KKgrav}$, providing an explanation for an early discovery of the graviton. In detail, this happens in an **RS model with intermediate branes,** where only gravity propagates in the whole extra dimension [35].

In any case, this new result forces us to reconsider our interpretation of T690 and opens an entirely new domain of exciting investigations.

The set of indications for X650 could include **two resonances,** almost degenerate in mass: a **tensor T690 decaying into WW/ZZ, h125h95, $\gamma\gamma$ and e+e- co-existing with a scalar H650 decaying into T350h125 and A450Z**.

**T$^{++}$450** observed into W$^+$W$^+$ could belong to an **isoquintuplet TQ of tensor resonances** comprising **T$^+$375**, indicated into W$^+$Z and **T350**, indicated into hh and b$\overline{b}$. This TQ satisfy **unitarity constraints** and therefore one assumes, by analogy, that **T690** could be the neutral component of the next TQ, while the charged components remain to be discovered.

This KK graviton interpretation naturally predicts the observation of **T690→$\gamma\gamma$** which is confirmed by ATLAS although with a **factor 100 suppression** with respect to ZZ. This suppression, interpreted in [35] for $\gamma\gamma$, could also be present for gluons and leptons, explaining the absence of a clear signal in lepton pairs and the dominance of WW/ZZ fusion with respect to ggF, but this remains to be understood.

CMS seems to bring an additional proof for **T690→γγ** using the photon tagging information provided by TOTEM.

**T690→e+e-** is indicated by ATLAS and CMS but does seem to fit in the RS bulk description which predicts a negligible BR.

With these RS reinterpretations, there is no need to invoke the Georgi Machacek model. The **scalar resonances** are accommodated into three Higgs isodoublets **3HD** [28]. This model is motivated since it allows **CPV** as well as natural flavour conservation[12]. It also allows to understand the dominant mode **H⁺130→b̄c** indicated by ATLAS in rare top decays.

The RS model suggests to interpret h95 as the **radion** predicted within extra dimension theories.

LHC results on the **triple Higgs coupling** $\kappa_\lambda$ provide a complementary approach to predict BSM heavy scalar resonances. They already **predict a scalar with mass below 1 TeV in most models**, with the notable exception of MSSM. H650 is a likely heavy scalar candidate which can significantly alter the Higgs potential, implying a measured value of $\kappa_\lambda$ **above the SM value**. The contribution from T350→hh could also affect the measurement of $\kappa_\lambda$.

Assuming RUN2+RUN3 data, we propose the following line of action:

- Confirm the tensorial interpretation of **T690→ZZ** with adequate selections
- Confirm **T690→γγ** inclusive and with TOTEM
- Confirm indications for **T690→e+e-** and look for T350→e+e-
- Confirm **T⁺⁺450→W⁺W⁺, T⁺375→ZW⁺ and T350→hh** and determine their spins
- Confirm ATLAS indication for **t→H⁺130 b→ b̄cb** and look for **T⁺⁺450→H⁺H⁺**
- Confirm ATLAS indication for **H⁺700→A450W⁺**
- Look for the **charged tensors, isospin partners of T690** in the neighbourhood of 700 GeV
- Look for the predicted **heavier KK graviton** at 1 TeV and 1.3 TeV
- Search for $Z_{KK}$ **and** $W_{KK}$ **gauge bosons** for masses circa **3 TeV**
- Reinforce the **phenomenology** of KK gravitons

The need for the later should be obvious after reading the present paper where, in this complex zoology, the concept of unitarity and interference effects seem to play a major role.

Needless to say that the various interpretations presented in this paper are moving targets, continuously subject to new experimental inputs. Our present feeling is that the **RS heavy graviton interpretation of T690** is consolidated by many, but not all, observations. Our present interpretation of the data inspired by the RS shows some disagreements with this model:

- Apparent absence of ggF production of T690
- Absence of decay of T690 into t t̄

---

12  The 2HD model with alignment is incompatible with CPV.

- Indication for doubly charged tensors, which insure unitarity for WW/ZZ, but are not predicted within the standard RS model

- Indication for decays of T690 into e+e-, not predicted within standard RS

One may therefore try to extract what seems to be the **less model dependent part**:

- The presence of a **narrow tensorial resonance which couples to e+e-** with formidable experimental prospects

- The likely presence of a doubly charged tensorial state, which allows to insure unitarity for WW/ZZ, predicting a light $H^+$ to allow for the decay $T^{++} \rightarrow H^+ H^+$

- The presence of BSM light and heavy scalars, requiring to go beyond the 2HD model

We hope that this note will trigger a common effort from ATLAS and CMS to test the tensor hypothesis in the various channels observed so far. Together with RUN2, **RUN3** should, by end of 2025, triple the total luminosity [46] sufficient, combining the two experiments, to confirm or dismiss most of the present indications at a 3 s.d. level significance.

**Acknowledgements:**
Contacts with **Yves Sirois** are gratefully acknowledged.
Remarks from **Heather Logan**, in particular on the H+130→$c\bar{b}$ indication, were very fruitful.
Essential contributions from **Juergen Reuter** to our understanding of the tensorial world are gratefully acknowledged with his patient and immediate answering to the questions of F.R.
Many thanks to **Raman Sundrum** for answering F.R. about the possibility of KK gravitons being lighter than KK vector mesons.
**Kaustubh Agashe** has provided essential and continuous inputs on the KK graviton scenario.
Many thanks to **Johannes Braathen** for enlightening explanations on the predictions for heavy scalar masses deduced from the measurements of $\kappa_\lambda$ performed at LHC.
Friendly and pertinent exchanges with **Andreas Crivellin** are gratefully acknowledged.
Very useful comments about the candidate CP-odd near $t\bar{t}$ threshold were provided by **Abdelhak Djouadi.**
Many thanks to **Christophe Royon** for patiently explaining the subtleties of operations of TOTEM with CMS.
F.R. wishes to thank **G. Unal** and **L. Fayard** for providing useful informations.

# APPENDIX  I

# Interpretation of t→H⁺130b→b̄cb within the Weinberg model

Reference [27] provides a detailed description of the Weinberg 3HD model [28] and its application to the evidence for t→H⁺130b→cb̄b provided by ATLAS. At its says, this evidence is weak and would not draw too much our attention if this particle was not implied by the low BR of H⁺⁺450→W⁺W⁺ and H⁺375→ ZW⁺, themselves implied by X60→ ZZ/WW. Appendix I is a simple summary of the results which can be found in [27]. The Weinberg model is characterized by the presence of two extra mixing angles. The three isodoublets have vacuum expectations v1, v2, and v3. One defines: $\tan\beta$=v2/v1 and $\tan\gamma$=sqrt(v1²+v2²)/v3.

There is an additional angle θ . *The parameter ϑ is a mixing angle between the two massive charged scalars and δ is a CP-violating phase.* Following [27], we take θ=−π/2.1 . The δ=0 as a starting point.

As usual the physical states are related to the EW eigenstates by a 33 matrix U:

$$\begin{pmatrix} s_\gamma c_\beta & s_\gamma s_\beta & c_\gamma \\ -c_\theta s_\beta e^{-i\delta} - s_\theta c_\gamma c_\beta & c_\theta c_\beta e^{-i\delta} - s_\theta c_\gamma s_\beta & s_\theta s_\gamma \\ s_\theta s_\beta e^{-i\delta} - c_\theta c_\gamma c_\beta & -s_\theta c_\beta e^{-i\delta} - c_\theta c_\gamma s_\beta & c_\theta s_\gamma \end{pmatrix} .$$

The decay widths depend on 3 quantities:

$$X = \frac{U_{d2}^\dagger}{U_{d1}^\dagger}, \qquad Y = -\frac{U_{u2}^\dagger}{U_{u1}^\dagger}, \qquad Z = \frac{U_{\ell2}^\dagger}{U_{\ell1}^\dagger} .$$

where the indexes u,d and ℓ are given in Table I.

|  | u | d | ℓ |
|---|---|---|---|
| 3HDM (Type I) | 2 | 2 | 2 |
| 3HDM (Type II) | 2 | 1 | 1 |
| 3HDM (Lepton-specific) | 2 | 2 | 1 |
| 3HDM (Flipped) | 2 | 1 | 2 |
| 3HDM (Democratic) | 2 | 1 | 3 |

*TABLE I: The five versions of the 3HDM with NFC and the corresponding u, d and ℓ values. Taking u = i means that the up-type quarks receive their mass from vi and likewise for d (downtype quarks) and ℓ (charged leptons).*

As in [27], we choose 3HDM (Flipped).

Mixing angles β and γ are constrained by the ATLAS indication for H⁺130 and the LHC searches for H⁺→τ⁺ν and H⁺→ c̄s+cb̄ and by the b→sγ measurement, as shown in figure 19.

We choose mH⁺₂=700 GeV, tanβ=15 and tanγ=10 and derive:
**v1=16.2 GeV v2=244 GeV v3=24.5 GeV**
**X=-1.05, Y=0.1 and Z=-0.1**

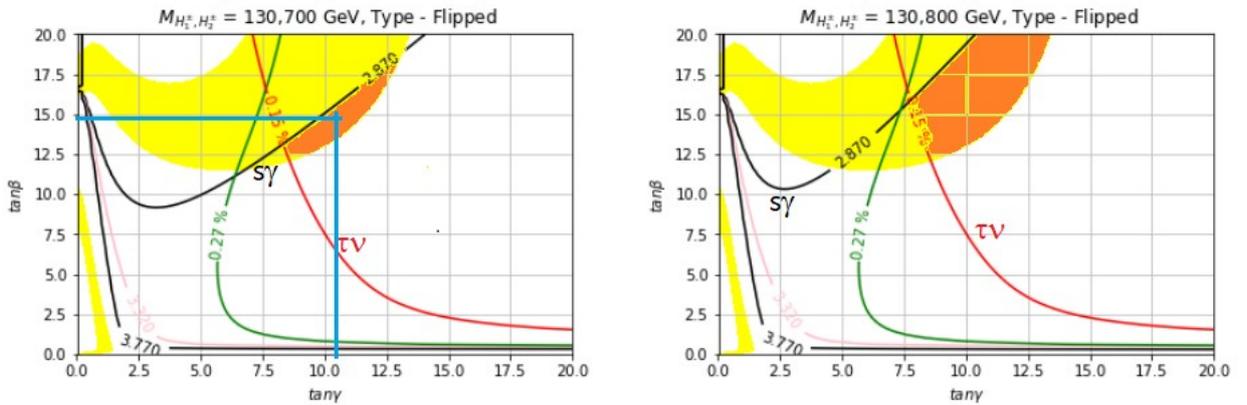

*Figure 19: The orange region represents the allowed regions for mH₁⁺=130 GeV and mH₂⁺=700 GeV (left panel) and mH₂⁺=800 GeV (right panel) in the 3HD (Flipped) model . The yellow area is given by the ATLAS indication for H⁺130. The black curves come from the b→sγ constraints, the red curve comes from LHC searches using H₁⁺→τ⁻ν , the green curve from c̄s+cb̄ channels. The chosen parameters are indicated in blue (left panel).*

We derive the various relevant widths from:

$$\Gamma(H^{\pm} \to \ell^{\pm}\nu) = \frac{G_F m_{H^{\pm}} m_{\ell}^2 |Z|^2}{4\pi\sqrt{2}} ,$$

$$\Gamma(t \to W^{\pm}b) = \frac{G_F m_t}{8\sqrt{2}\pi}[m_t^2 + 2m_W^2][1 - m_W^2/m_t^2]^2 ,$$

$$\Gamma(H^{\pm} \to ud) = \frac{3G_F V_{ud}^2 m_{H^{\pm}} (m_d^2 |X|^2 + m_u^2 |Y|^2)}{4\pi\sqrt{2}} .$$

$$\Gamma(t \to H^{\pm}b) = \frac{G_F m_t}{8\sqrt{2}\pi}[m_t^2 |Y|^2 + m_b^2 |X|^2][1 - m_{H^{\pm}}^2/m_t^2]^2 .$$

*which gives us : **BR($\tau^+\nu$)=18 % BR($c\bar{s}$)=22%   BR($c\bar{b}$)=60 %    BR(t→H$^+$130b)=0.24 %***

*One accordingly understands the formidable challenge for LHC to extract H$^+$130 from top decays. The dipole moment of the neutron gives an upper bound |sin$\delta$|<0.9. Even if we take the largest allowed CPV, this will not change drastically our conclusions on H$^+$130 branching ratios.*

e$^+$e$^-$ should provide a very clean situation by using the channel **e$^+$e$^-$→ H$^+$H$^-$** with the presence of up to **2b quarks**, which allows a separation from W$^+$W$^-$,and by operating below the top threshold.

e+e- should also provide a powerful test of CPV in the Weinberg model [29] through the measurement of e+e-→ Z→ hihj where hi are the five neutral bosons of this model. This process gives zero for h95h125 in case there is CP conservation but can significantly differ from zero otherwise. Assuming that h2 is the SM scalar and that h1 is h95 and assuming that there is alignment such that h2 behaves as the SM, figure 20 gives the degree of Z affinity between the 5 neutral states[13]. It shows that the coupling of the SM h2 couples weakly to the 4 other scalars as a consequence of the SM behaviour of h2. In contrast one expects a large coupling for e+e-→ h95hA152, which is good news for circular machines.

 It indicates that the CP odd states A152 and A450 contain a CP even component such that ZA152A4320 is non zero.

H$^+$700 can decay into A450W$^+$→t$\bar{t}$W$^+$ which has presumably resulted in its observation in an inclusive jet-jet mode requesting a large pt lepton + missing energy [17]

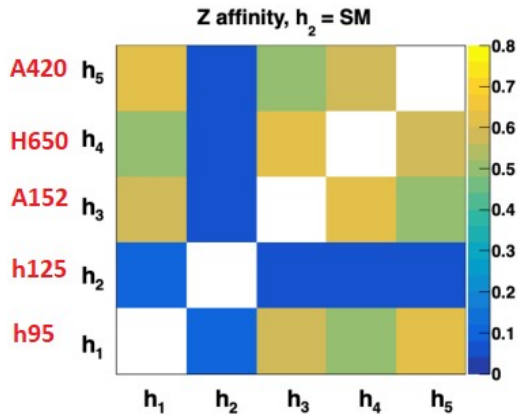

*Figure 20: Average Z affinity of states hi and hj neutral scalars assuming h2 is the SM scalar.*

---

# APPENDIX  II

## Crude evaluations for T690, T350 and T$^{++}$450 WW widths

From the W$^+$W$^-$ analysis, CMS concludes that VBF is dominant and measures VBFBR(W$^+$W$^-$)=160±50 fb. In a tensor scenario, one expects VBFBR(ZZ)=80±25 fb which predicts 56 events times the efficiency for the four lepton channel. Figure 2 indicates an excess of 10 events which corresponds to a reconstruction efficiency of 20%, which seems reasonable taking into the tagging efficiency. In figure 1, the excess is larger but compatible with a small contribution from ggF, taking into account the tagging efficiency. For a tensor, one predicts $\Gamma$690$_{WW}$=14 GeV and $\Gamma$690$_{ZZ}$=7 GeV, compatible, within errors, with the total width found taking into account interference effects.

From these values one can extract **c=κ/$\overline{MPL}$** from [22]:     $\Gamma\left(G \to W_L^+ W_L^-\right) \;\approx\; \dfrac{(c\,x_n^G)^2\, m_n^G}{480\pi}$

Assuming that T690 is the **second KK recurrence,** $x^G_2$=7,  one finds **c=0.7**.

The two photon measurement from ATLAS gives $\Gamma$690$_{\gamma\gamma}$ =70 MeV.  It is too early to decide on $\Gamma$690$_{ee}$.

Figure 7 shows the predicted RS cross sections given by CMS. Taken at face value, they would exclude KK gravitons up to 1.8 TeV.  This result assumes a large ggF cross section while present results seem compatible with a dominant VBF contribution.

The RS model also allows predicting the width of T350 into W$^+$W$^-$ using above formula replacing the value of $x^G_1$ by 3.8, which gives $\Gamma$350$_{W+W-}$=2GeV.

 To predict the width of T$^{++}$450 into W$^+$W$^+$ one can assume that the coupling constant is the same as for T350 since the two resonances belong to the same isomultiplet which gives $\Gamma$450$_{W+W+}$=1.3GeV taking into account the statistical factor for identical bosons. From the measured cross section of ATLAS one can deduce that BR(W+W+)~20%, which requires additional decay modes of the type **H$^+$W$^+$ and H$^+$H$^+$**.

# APPENDIX III

# KK gravitons couplings

## Coupling of a graviton to e+e- and γγ

The coupling of a standard graviton reads:

$$\mathscr{L}_{V,f}^G = \frac{k_{V,f}}{\Lambda} \, T_{\mu\nu}^{V,f} \, G^{\mu\nu}$$

where, a-priori, one allows for non-universal couplings for fermions and bosons. This is evidently the case for the KK gravitons where the coupling depends on the localization in the bulk of the various particles.

For instance, the expression for the electromagnetic part reads [44]:

$$T^{\mu\nu} = T_e^{\mu\nu} + T_\gamma^{\mu\nu}, \;\; \text{with}$$

$$T_e^{\mu\nu} = \bar{\psi}\gamma^\mu \frac{i\overleftrightarrow{D}^\nu}{2}\psi, \qquad T_\gamma^{\mu\nu} = -F^{\mu\alpha}F^\nu{}_\alpha + \frac{g^{\mu\nu}}{4}F^2$$

which illustrates how electrons and photons can couple to the graviton.

## Unitarity sum rules for the tensor isoquintuplets

The GHW sum rule applied to WW→WW for the Higgs sector allows to write:

$$g^2(4m_W^2 - 3m_Z^2 c_W^2) \stackrel{\rho \simeq 1}{\simeq} g^2 m_W^2 = \sum_k g_{W^+W^- H_k^0}^2 - \sum_l g_{W^+W^+ H_l^{--}}^2$$

One can generalize this rule to include the tensors. For tensors, there is no SM contribution to the left hand-side, while the right-hand receives the following contribution:

$$\Sigma_i \left( g^2{}_{W+W-GKKi} - g^2{}_{W+W+GKKi++} \right)$$

Note that what only matters are the **coupling constants to WW** and not the masses nor the BR of the gravitons. If each KKi recurrence of gravitons belongs to an isoquintuplet, couplings of GKKi to W+W- and of GKKi++ to W+W+ are equal and one predicts **cancellation** and therefore a negligible contribution from KK gravitons to the GHW sum rules.

Apparently this does not solve the issue for vector and scalar KK contributions to the sum rule which, however, will have reduced impact if these KK states are much heavier than the KK gravitons, in which case unitarity will be violated at a much larger energy.